# Self-intercalation as origin of high-temperature ferromagnetism in epitaxially grown $Fe_5GeTe_2$ thin films

M. Silinskas[1], S. Senz[1], P. Gargiani[2], A.M. Ruiz[3], B. Kalkofen[1], I. Kostanovskiy[1], K. Mohseni[1], J.J. Baldoví[3], H. L. Meyerheim[1], S.S.P. Parkin[1] and A. Bedoya-Pinto*[3]

[1] NISE Department, Max Planck Institute of Microstructure Physics, Halle, Germany
[2] ALBA Synchrotron Light Source, Barcelona, Spain
[3] Institute of Molecular Science, University of Valencia, Spain
*Correspondence to: amilcar.bedoya@uv.es

## Abstract

The role of self-intercalation in 2D van der Waals materials is key to the understanding of many of their properties. Here we show that the magnetic ordering temperature of thin films of the 2D ferromagnet $Fe_5GeTe_2$ is substantially increased by self-intercalated Fe that resides in the van der Waals gaps. The epitaxial films were prepared by molecular beam epitaxy and their magnetic properties explored by element-specific x-ray magnetic circular dichroism that showed ferromagnetic ordering up to 375 K. Both surface and bulk sensitive x-ray absorption modes were used to confirm that the magnetic signal is of an intrinsic nature. Fe occupation within the van der Waals gap was determined by x-ray diffraction which showed a notably higher occupation with respect to bulk crystals. We thus infer, supported by first-principles calculations, that the higher magnetic ordering temperature results from an increased exchange interaction between the individual $Fe_5GeTe_2$ layers mediated by Fe atoms residing within the van der Waals gaps. Our findings establish self-intercalation during epitaxial growth as an efficient mechanism to achieve high-temperature magnetism in a broad class of van der Waals materials.

## Main

The occurrence of long-range magnetic order in two-dimensional systems has emerged as an intriguing topic, as the first experimental observation, achieved by the successful exfoliation of atomically-thin flakes from bulk crystals [1-3], challenged the long-standing Hohenberg-Mermin-Wagner Theorem [4]. It has been then established that a sizable magnetic anisotropy is needed to stabilize long-range magnetic order in the strictly 2D limit, either by virtue of an out-of plane uniaxial anisotropy [1,3] or an easy-plane anisotropy [5]. While out-of plane magnetized 2D systems can be potentially exploited for domain wall motion and spin-torque applications [6], the limiting factor is still their low magnetic ordering temperature (see [7] for a recent review).

To this end, efforts to achieve high-temperature ferromagnetism in layered magnets are currently being pursued. In semiconducting magnets such as $Cr_2Ge_2Te_6$, strong charge doping via ionic liquid gating has been shown to drastically increase the Curie Temperature ($T_c$) [8]. On the other hand, in metallic systems such as $Fe_3GeTe_2$ (F3GT), where electric gating is less efficient [9], other approaches have been followed, such as tuning the Fe concentration within the crystal structure [10-11], Cobalt co-doping [12], high pressure [13], and proximity effects with a topological insulator [14]. An increase in the Fe-composition (from x=3 to x=5) in $Fe_xGeTe_2$ crystals has shown to boost $T_c$ from 240 K up to 310 K [10,11], whereas a partial substitution with Co-atoms (up to 26%) enhances $T_c$ up to 328 K [12]. Magnetic ordering up to 360K has been observed after pressure loads of 14 GPa in diamond anvil cells [13]. In thin film samples, e.g. in F3GT/$Bi_2Te_3$ heterostructures, it has been argued that the proximity to the topological insulator induces a $T_c$ enhancement up to 380 K [14]. Most recently, a record high-temperature magnetic ordering temperature (>350K) in epitaxially grown $Fe_{5-x}GeTe_2$ thin films were reported by two independent groups [15,16] in the *as-grown* state (without co-doping, chemical gating nor under pressure). These intriguing results are not fully understood, especially whether effects such as strain, inhomogeneous Fe-distribution, or interface effects play a role in such a dramatic enhancement.

In this work, we unveil the origin of the record-high Curie Temperature in epitaxial $Fe_5GeTe_2$ films combining detailed x-ray diffraction (XRD), x-ray magnetic circular dichroism (XMCD) and first-principles calculations. Going beyond previous experimental and theoretical intercalation studies of the parent compound $Fe_3GeTe_2$ [17], our results allow to establish a clear-cut relation between the structural self-intercalation of the van der Waals (vdW) gaps by Fe atoms and intrinsic high-temperature ferromagnetism with a $T_C$ of 380 K. We attribute the substantially higher magnetic ordering temperature of our MBE-grown $Fe_5GeTe_2$ films -as compared to bulk crystals- to an enhancement of the *interlayer* exchange interaction between the individual F5GT sheets mediated by Fe atoms residing within the vdW gaps. Our findings highlight the importance of the combination of a highly precise atomically-resolved structural and magnetic characterization and establish a promising pathway to prepare robust high-$T_C$ 2D systems for spintronic applications. The strategy to boost $T_c$ via self-intercalation relies on growth kinetics and is widely applicable to any layered magnet that can be synthesized with a low-rate, low temperature growth method.

$Fe_5GeTe_2$ (F5GT) epitaxial films were grown by molecular-beam epitaxy on $Al_2O_3$ (0001) substrates and capped with a crystalline tellurium layer prior to ex-situ characterization. A schematic model of the F5GT crystal structure in cross-sectional view is presented in Fig. 1a, underlining the two van der Waals gaps separating the three F5GT sheets within the unit cell. Standard x-ray diffraction was first employed to analyze the lattice metric of the films, typical $\theta - 2\theta$ scans are presented in Fig. 1b. Besides the strong reflection at $2\theta = 41.6°$ arising from the $Al_2O_3$ (0 0 1) substrate, the peaks correspond to the (0 0 3), (0 0 6) and (0 0 9) reflections of hexagonal $Fe_{4.87}GeTe_2$ [10] (space group R3m1 (#160)). The inferred c lattice constant of the films is equal to 28.75 Å (sample 1, thickness d=38nm) and 28.60 Å (sample 2, thickness d=18nm), in agreement with those derived by cross-sectional transmission electron microscopy (TEM) (Fig. 1c), from which a spacing of 9.56 Å between F5GT sheets is found (c=28.7 Å). The observed deviation from the bulk crystal values (c=29.25 Å in Ref. [10]) suggests a compression of the c-axis that arises during epitaxial growth. The TEM image in Figure 1c also shows an enlarged view of two sections of the film with different contrast. They correspond to twisted structural domains, i.e. (1 1 0) and (1 1 1) orientations which have both an interplanar distance of 2.04 Å (lattice constant a = b = 4.08 Å). Overall, the results are consistent with a single-phase F5GT film with slightly different lattice parameters (shorter in *c* and larger in *a, b*) as compared to the bulk crystals [10].

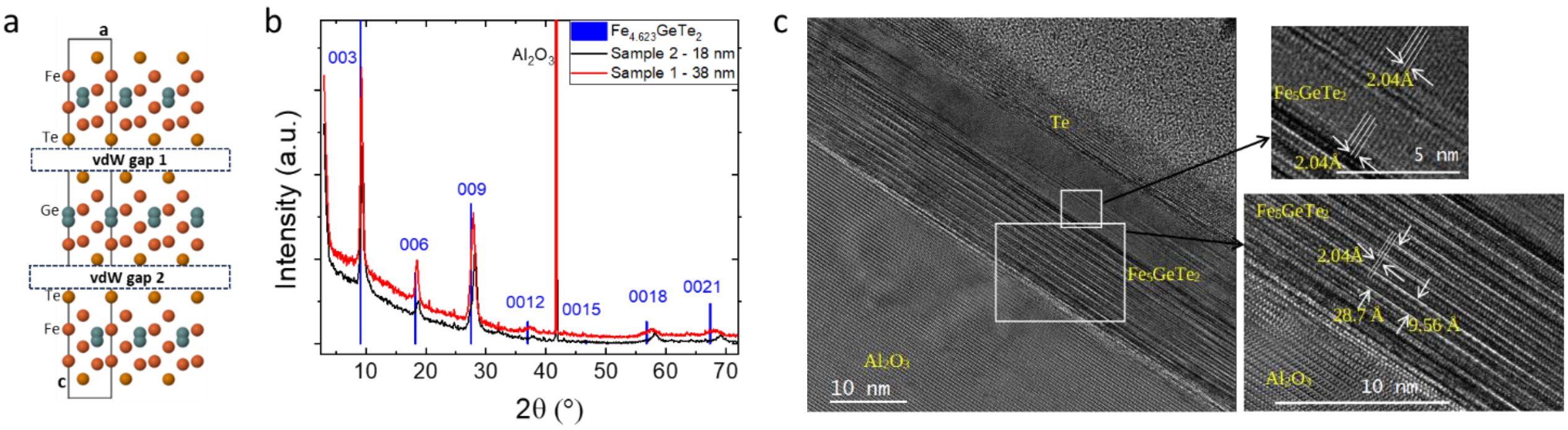

***Figure 1**. a) Model of the crystal structure of $Fe_5GeTe_2$ (F5GT) in cross-sectional view, highlighting the presence of two van der Waals gaps within each unit cell. b). Specular Theta-2Theta scan along the (00L) direction in reciprocal space of the F5GT films, showing only the expected reflections of the $Fe_{5-x}GeTe_2$ phase and of the $Al_2O_3$ substrate. c) TEM micrograph of a F5GT film with tellurium capping layer (Sample 2 – 18 nm). A close-up image of two representative regions of the F5GT film are shown, along with the measured interplanar distances. The scale was calibrated by using the (0 0 1) and (1 1 0) interplanar distances of the $Al_2O_3$ substrate.*

The magnetic properties of the F5GT thin films were investigated via x-ray magnetic circular dichroism (XMCD), a technique which is reliable to determine the intrinsic magnetic properties of ultrathin magnets down to the monolayer and sub-monolayer limit [5,18], as well as of cleaved layered crystals [19]. The XMCD contrast results from the difference in absorption when exciting with opposite X-ray photon helicities, -here recorded in the vicinity of the Fe $L_{2,3}$ absorption edges- and is proportional to the magnetic moment (in $\mu_B$/atom) of the atomic species probed. Figure 2a shows the field-dependent XMCD intensity at the Fe $L_{2,3}$ edge (sample 1) at 300K, both in normal (NI) and grazing incidence (GI, 70 deg off-normal), which corresponds to an out-of-plane and an almost in-plane magnetic field alignment, respectively. A clear in-plane easy magnetization axis is evidenced by a square-loop hysteresis with high remanence and a low coercive field (50 mT) in the case of GI, while a typical hard-axis magnetization loop is observed for NI (extracted anisotropy field: 1T). The full XMCD spectra at the Fe $L_{2,3}$ edges recorded at B=0 is presented in Figure 2b: the remarkable difference of the remanent magnetization under GI and NI is directly visualized in the raw energy scans. The Fe $L_{2,3}$ lineshape has two contributions, one (707.6 eV) showing XMCD contrast -ascribed to Fe atoms within the F5GT structure- and the other (708.9 eV) being magnetically inactive, belonging to a thin region of oxidized Fe at the F5GT surface (1 to 1.5 monolayers) owing to a finite permeability of the Te-capping layer. Considering a Te-capping layer thickness of 7 nm (TEM image in Fig. 1c), the top F5GT layers are still within the electron escape depth to allow for XMCD detection. More details about the Fe $L_{2,3}$ lineshape and XMCD contrast (comparing bulk vs surface measurements) are discussed in the Supplementary Information (Figs. S4-S5).

The quality of the XMCD spectra allows to perform a quantitative sum-rule analysis [20] and extract the orbital and the effective spin moments at the Fe-sites, which amount to $m_L$= 0.14 μB and $m_{S_eff}$= 1.48 μB, the latter being estimated as a lower bound (see Supp. Information and Fig. S6 for more details). The averaged magnetic moment per Fe-site (1.62 μB at 300 K) would then fall in the same range as the values reported in bulk crystals [10,19,21] and epitaxial films [22]. Figure 2c displays the temperature-dependent XMCD signal collected at remanence (B=0), scaled to the remnant atomic moment ($m_r$) per Fe-sites (via sum rules) to account for the different film thickness of sample 1 and 2 (38 and 18 nm, respectively). Both samples follow the same trend, i.e. the magnetic moment at the Fe-site continuously decreases with temperature and a $T_c$ = 380K can be extrapolated. The corresponding raw XMCD spectra are shown in the inset of Figure 2c, showing a non-zero XMCD signal up to 375K at both Fe $L_3$ and $L_2$ edges. In order to resolve the low-field region of the field-dependent magnetization, the XMCD detection scheme is changed from total electron yield (via drain current) to fluorescence yield, that relies on a silicon drift diode that is not prone to signal artifacts due to magnetic field switching around 0T (a comparison between both detection methods can be found in the Supp. Note 2). The temperature evolution of the field-dependent XMCD signal using fluorescence yield is shown in Figure 2d. Clear hysteresis loops up to 375K could be well resolved, demonstrating a robust ferromagnetic ordering persisting well-above room-temperature. The occurrence of high-temperature ferromagnetism has been reported in previous studies of F5GT thin films [15,16], but not assessed with an element-specific technique (XMCD) to such extent. Importantly, unlike the previous studies where secondary phases such as $FeTe_2$ [16] and $Fe_3GeTe_2$ [15] have been found within the F5GT films, the samples presented here are single-phase such that all the magnetic signals stemming from the Fe $L_{2,3}$ edge are intrinsic to the F5GT structure. The inferred anisotropy field (1T) is a factor of 10 higher than in previous reports (about 0.1 T in [15]), underlining the robustness of the ferromagnetic behavior in our samples. Moreover, by carrying out both total electron yield (surface sensitive, with a probing depth of the order of a 3-5nm after accounting for the capping layer thickness) and fluorescence yield (bulk sensitive) measurements (Fig. 2c and 2d), we conclude that the high-temperature ferromagnetism is a property that emerges homogeneously across the film thickness and does not arise from a surface/interface layer or any proximity interaction with the substrate.

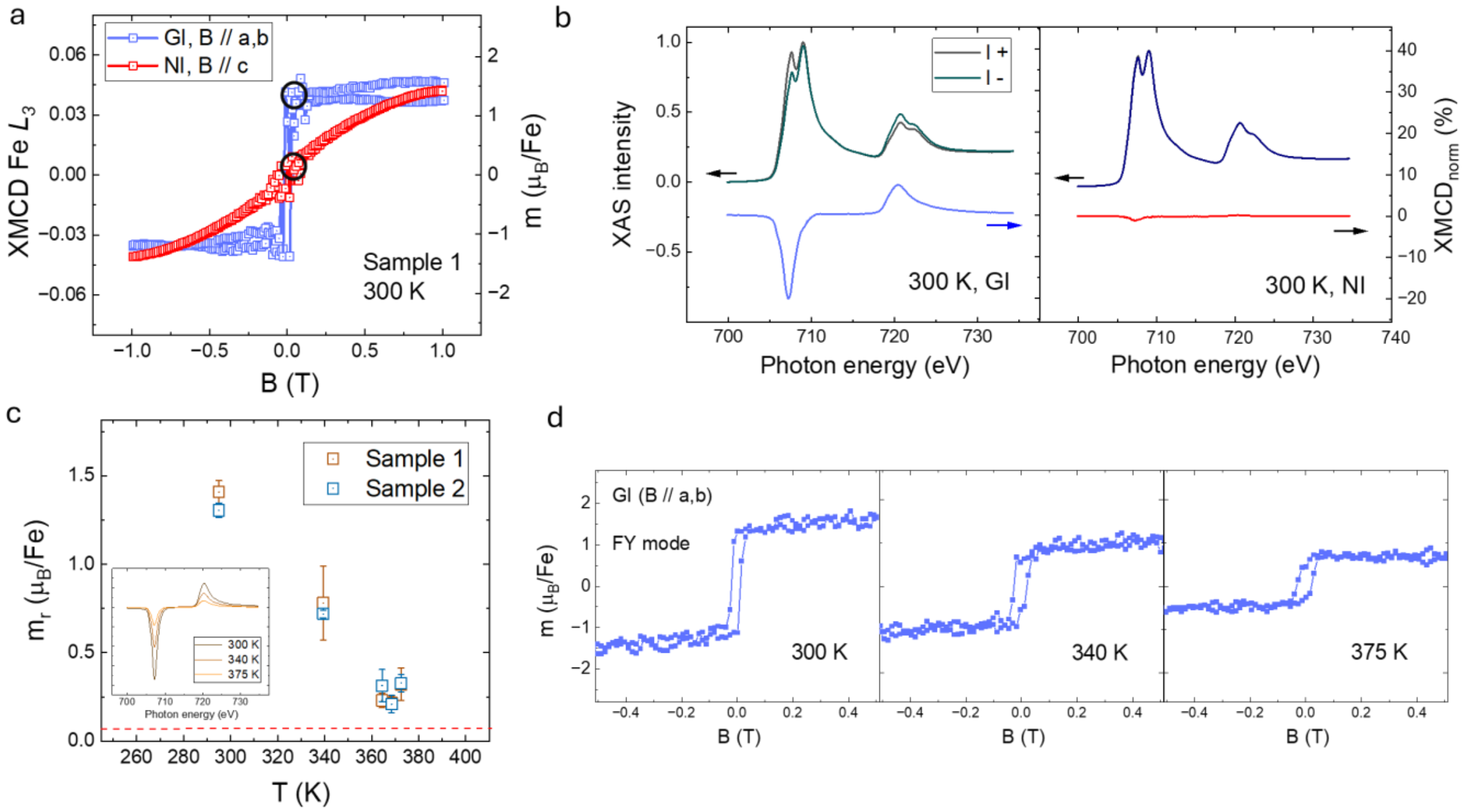

***Figure 2.*** *a) Normalized XMCD intensity at the Fe $L_3$ edge as a function of magnetic field, taken in grazing (B in-plane) and normal incidence (B out-of-plane) geometry (Sample 1). The remanence is highlighted by circles and the corresponding spectra is shown in panel b) for each orientation. The data is acquired at 300K by total electron yield (TEY). b) Helicity dependent X-ray absorption spectra at B=0, featuring a large XMCD contrast (20%) at grazing incidence, while at normal incidence the XMCD contrast is barely detectable, consistent with a sizable magnetic anisotropy. c) Temperature dependence of the remnant magnetic moment $m_r$ (XMCD at B=0), showing a detectable signal up to 375 K in both samples (d=18nm / d=38nm). The dashed red line represents the measurement resolution limit. Inset: Corresponding XMCD spectra across the Fe $L_3$ and $L_2$ edge at various temperatures. d) XMCD hysteresis loops in grazing incidence using fluorescence (FY) detection mode to resolve the low field region.*

In order to unveil the origin of the high Curie-Temperature obtained in the epitaxial films with respect to the bulk crystals (290-300 K), we have performed highly precise x-ray diffraction measurements (for technical details and extended discussion see Supp. Info; for similar experiments of layered materials and ultra-thin films we refer to our previous work [23,24]) Fig. 3a shows a schematic of the structural model which is based on best fit given by a goodness of fit (GOF) [25] of 0.91. The atoms are labelled from (1) to (7) which refers to Table S1 in the Supplementary Information. We find that the film structure closely resembles that of single crystalline bulk $Fe_{5-x}GeTe_2$ [26]. In general, there is a strong vertical disorder, which is especially pronounced at Ge(1) and Fe(5), for Fe(5) it was treated by using split sites (5 and 5s) each occupied by a site occupancy factor (SOF) of 50%. The structural refinement also indicates the presence of vacancies at several Fe sites, and most importantly, a substantial amount of Fe within the vdW-gap between the F5GT sheets. In Fig. 3a and 3b these atoms are indicated by small pink balls labelled by Fe(6) and Fe(7), respectively. The detailed quantification is outlined in Fig. 3c which shows the contour plot of GOF versus the SOF's related to these atoms. The GOF minimum is found for Fe located in Wyckoff sites 18c and 3a in space group R3m (Nr. 160 in Ref [27]). These are occupied by about 8 and 20%, respectively. Notably, if the vdW site occupancy is neglected, a substantially higher GOF of 1.3 is obtained, i.e. the consideration of the Fe atoms in the vdW gaps is mandatory to achieve a satisfying fit. The overall stoichiometry without consideration of the vdW-Fe atoms is equal to $Fe_{4.77}GeTe_{1.90}$ (for the F5GT single crystal we obtain $Fe_{4.52}GeTe_{1.90}$, i.e. the vacancy concentration is higher than in the film). However, the consideration of vdW-site occupancy by Fe atoms and preserving the R3m symmetry leads to $Fe_{5.54}GeTe_{1.90}$. We estimate the uncertainty of the SOF determination to lie in the range of 0.02-0.05 range so that an approximate uncertainty for the total Fe concentration of ±0.15 is derived. We compare the results for the film with those found for the single crystal, where vdW-site Fe atoms are also identified, albeit by a less concentration (18c: SOF=0.04, 3a: SOF=0.17) – the inferred stoichiometry is $Fe_{4.94}GeTe_{1.90}$. It is worth noting that the stoichiometry estimate by site occupancy factors in x-ray diffraction is sensitive only to those atoms that are crystallized in the F5GT structure, providing a more accurate, phase-selected model than other standard composition determination methods (EDX, RBS) which rely on volume integration. The fact that the Fe surplus is being incorporated in the vdW gap sites is confirmed by a careful inspection of the TEM images (Figure 3d), where atomic contrasts in the gap between the bright Te(1) and Te(2) atoms can be detected (green boxes). In particular, the two Wyckoff positions for in-gap Fe according to the structure in Figure 3a can be distinguished: one in the middle of the gap (Fe 7) and the other closer to the upper Te(2) atomic row (Fe 6).

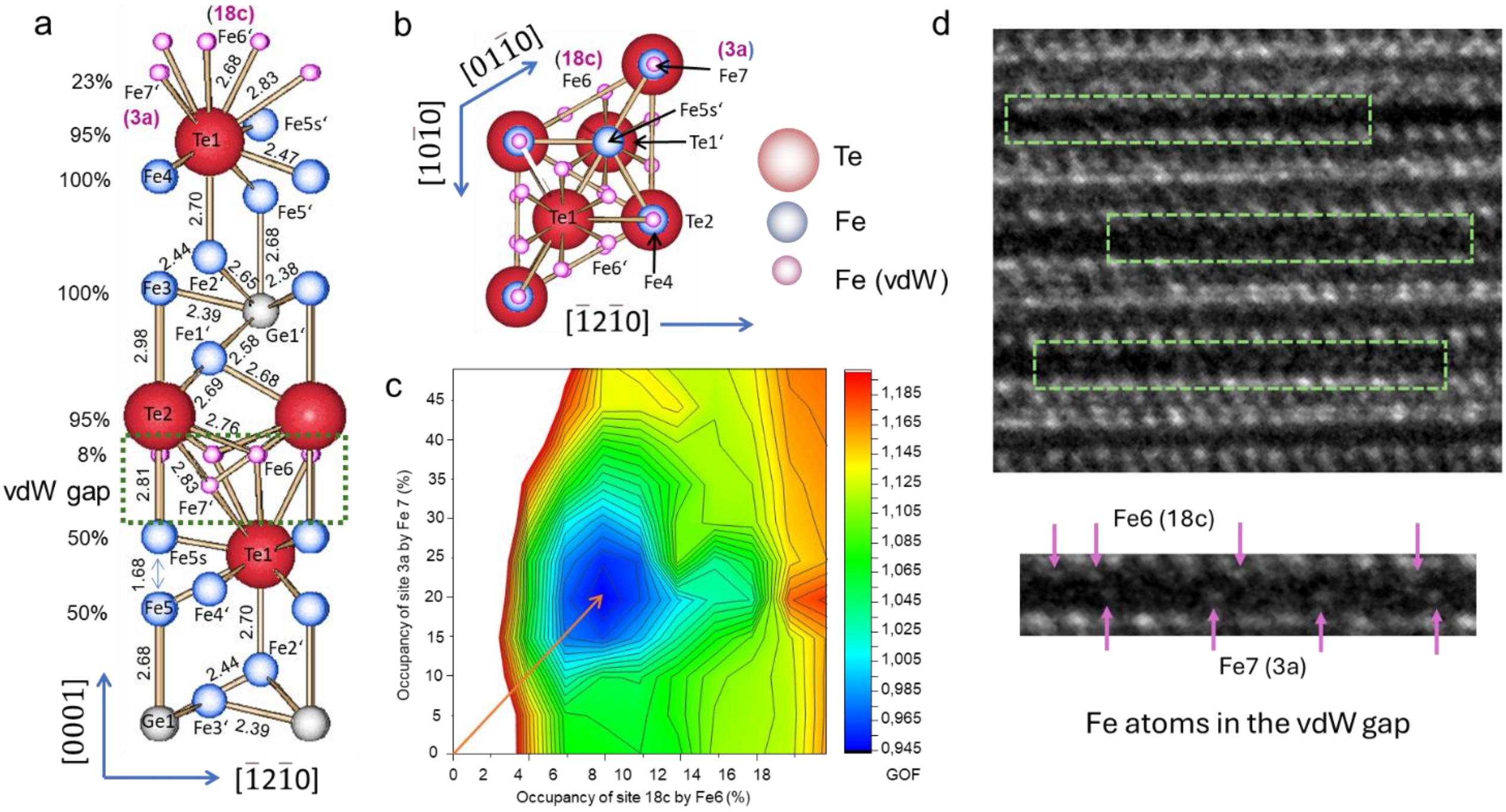

**Fig. 3.** *(a): Structural model of the F5GT film viewed in projection along the a-axis. For clarity, the unit cell is only shown up to z=0.55 lattice units along the c-axis. Atom labels refer to Table S1 of the Supplementary Information, those with primed numbers represent symmetry equivalent atoms. Interatomic distances are given in Ångstrøm units, the site occupancy factors in percent are given on the left. The location of van der Waals gaps are marked with dashed boxes. (b) Corresponding top view emphasizing the coordination of the Te atoms by Fe atoms (6) within the vdW gaps. (c): Contour plot of GOF versus SOF of Fe(6) and Fe(7) which occupy vdW-type sites at site 18c and 3a in SGR 160 (R3m), respectively. The arrow emphasizes the difference between the best fit at the GOF minimum (0.94), whereas the condition without any vdW-site occupancy leads to GOF ≈ 1.30 (white area). (d) Detailed view of a transmission electron microscopy image, highlighting the presence of Fe-atoms in the vdW gaps in the two different sites Fe(6) and Fe(7), in agreement with the x-ray diffraction refinement analysis.*

This noticeable additional amount of Fe within the structure ($Fe_{5.51}$ vs $Fe_{4.94}$) found by XRD and corroborated by transmission electron microscopy has important implications for the magnetic properties of the F5GT films. The van der Waals sites in the structure (Fe6 and Fe7) act as linkers between neighboring F5GT sheets along c-direction, thus being crucial for the strength of the interlayer magnetic exchange interaction. However, this interaction is also affected by other parameters such as the distance between the atomic planes along c-direction, which in our films is 0.3-0.5 Å shorter with respect to the bulk structure. In order to discern these concomitant effects, we have carried out density-functional theory (DFT) calculations using our experimentally obtained structural parameters for F5GT epitaxial films (Table S1), and computed the exchange interactions $J_{ij}$, the magnetic moments on each Fe-site and the resulting $T_C$ (for details see Supplementary Note 6). While merely changing the c lattice parameter leads to a very small variation of $T_c$, the specific inclusion of Fe in the vdW gaps, i.e. Fe(6) sites (Wyckoff 18c) leads to an large enhancement of $T_c$ – by a factor of 1.5– (see Figure S14), being the interlayer exchange coupling the decisive factor (see $J_{ex}$ values from Tables S4 and S5). Hence, we attribute the experimentally obtained $T_c$ of 380 K (compared to 293K – see Ref. [22]) to a higher occupation of Fe-atoms especially at the Fe (6) sites (Wyckoff 18c) which has the highest multiplicity in the structure of $R3m$ symmetry. The occupation of Fe-atoms in vdW sites and the associated enhancement of the interlayer exchange interaction is in line with the observations of a high $T_c$ using XMCD fluorescence yield mode, which probes the magnetism of the Fe-atoms throughout the whole film thickness. The driving force of such noticeable $T_c$ enhancement lies, however, on the crystal growth conditions. The synthesis of the bulk crystals (high temperature reaction around 700 ºC and the quenching of it [10]) is likely to produce more Fe-vacancies in the lattice than the epitaxial films which are grown at much lower temperatures (430 ºC). Additionally, the lower growth rate (about 10nm/h) employed in our epitaxial, layer-by-layer growth provides sufficient time for the Fe- vdW sites to populate more efficiently, being diffusion also facilitated in the nanometer thickness range. It is worth to recall that increasing the composition of Fe up to $Fe_6GeTe_2$ does not alter the magnetic properties of the bulk crystals [28], while modifying the cooling rate during synthesis [29] does have an impact on the Curie-Temperature (albeit only within ΔT=5-10 K). This underlines the role of growth-assisted diffusion, whereby populating specific crystal sites is shown to be more efficient than just increasing the Fe-composition. In this context, our experimental evidence of self-intercalation during epitaxial growth is of key importance to interpret other independent studies that also employed molecular-beam epitaxy for F5GT growth and observe a drastic enhancement of $T_c$ well above room-temperature [15,16], except when high-temperature post-growth annealing processes are carried out [22]. We also stress that self-intercalation processes inherent to low-temperature, low-rate growth (e.g. molecular-beam epitaxy) are present in other parent compounds such as $Fe_3GeTe_2$ which exhibit a record-high $T_c$ in the thin film limit [14]. The far-reaching implication of our combined x-ray magnetic dichroism and diffraction results is that adjusting growth kinetics is shown to be an efficient way to impact the occupation of specific Fe-sites and hence tailor magnetic parameters such as exchange interaction and magnetic anisotropy. We anticipate additional effects of Fe-intercalation such as modification of magnetic pinning, leading to complex spin textures and tunable domain-wall motion; as well as a higher resistivity which can be useful for 2D magnet-based spintronic heterostructures, i.e. to achieve efficient electrical switching of the magnetization via spin-orbit torques. The demonstration of self-intercalation during epitaxial growth, leading to intrinsic high-temperature ferromagnetic order, thus constitutes a significant advance for understanding and developing a new generation of high-$T_c$ layered magnets for spintronic device applications.

**Acknowledgments**

A.B.-P. and P.G. thanks CELLS-ALBA for the allocation of synchrotron radiation beamtime under granted proposals 2022097135 and 2022025755. A.B.-P. acknowledges support from the Generalitat Valenciana (grant CIDEGENT/2021/005) and from the Spanish MCIU (grant PID2023-149494NB-C31). J.J.B. acknowledges the European Union (ERC-2021-StG101042680 2D-SMARTiES) and the Generalitat Valenciana (grant CIDEXG/2023/1). A.M.R. thanks the Spanish MIU (Grant No FPU21/04195). S.S.P acknowledges funding by the Deutsche Forschungsgemeinschaft (DFG, German Research Foundation) – project no. 471731263; and partial funding by the European Union (FUNLAYERS, project number 101079184). Views and opinions expressed are however those of the authors only and do not necessarily reflect those of the European Union. Neither the European Union nor the granting authority can be held responsible for them.